\documentclass[10pt,prl,aps,twocolumn,showpacs]{revtex4}
\usepackage{graphicx}

\begin{document}

\title{Multi-critical point in a diluted bilayer Heisenberg 
quantum antiferromagnet}

\author{Anders W. Sandvik}
\affiliation{Department of Physics, {\AA}bo Akademi University, 
Porthansgatan 3, FIN-20500 Turku, Finland}

\date{\today}


\begin{abstract}
The $S=1/2$ Heisenberg bilayer antiferromagnet with randomly removed 
inter-layer dimers is studied using quantum Monte Carlo simulations. 
A zero-temperature multi-critical point $(p^*,g^*)$ at the classical 
percolation density $p=p^*$ and inter-layer coupling $g^* \approx 0.16$ 
is demonstrated. The quantum critical exponents of the percolating cluster 
are determined using finite-size scaling. It is argued that the associated 
finite-temperature quantum critical regime extends to zero inter-layer 
coupling and could be relevant for antiferromagnetic cuprates doped with 
non-magnetic impurities.
\end{abstract}

\maketitle

\begin{figure}
\includegraphics[width=4.5cm]{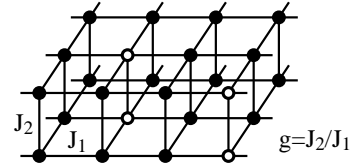}
\caption{The bilayer with intra- and inter-plane couplings 
$J_1{\bf S}_i \cdot {\bf S}_j$ and $J_2{\bf S}_i \cdot {\bf S}_k$. 
Solid circles represent magnetic sites ($S=1/2$). Two removed dimers are 
indicated by open circles.}
\label{fig:bilayer}
\end{figure}

\begin{figure}
\includegraphics[width=5.5cm]{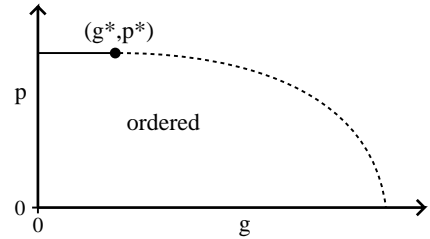}
\caption{Schematic $T=0$ phase diagram for the Heisenberg bilayer with 
coupling $g$ and a fraction $p$ of the inter-plane dimers removed. Percolation 
and quantum phase transitions are indicated by the solid horizontal line 
and dashed curve, respectively. The circle indicates the multi-critical 
point.} \label{fig:phasediag}
\end{figure}

Randomly diluted quantum spin systems combine aspects of the percolation 
problem \cite{stauffer} with the physics of thermal and quantum fluctuations. 
In systems that can be tuned through a $T=0$ phase transition as a function of
some parameter one can hence study divergent quantum fluctuations coexisting 
with classical fluctuations due to percolation. A multi-critical point, where 
the two types of fluctuations diverge simultaneously, is realized in the 
transverse Ising model with dimensionality $D > 1$ \cite{harris,senthil}. In 
models with $O(N)$ symmetry and $N > 2$ such a point was believed not to 
exist, because quantum fluctuations were argued to always destroy the 
long-range order on the percolating cluster \cite{senthil}. Several studies 
of diluted 2D Heisenberg antiferromagnets were consistent with this scenario 
\cite{wan,behre,chen}. However, recent quantum Monte Carlo simulations have 
shown that long-range order in the 2D Heisenberg model persists until the 
percolation point \cite{kato,perc1,perc2} and that the percolating cluster 
is ordered as well \cite{perc1,perc2}. This implies that the phase transition
is a classical percolation transition. It also suggests that a multi-critical 
point, at which the percolating cluster is quantum critical, could be reached 
by including other interactions. In this Letter it will be shown that the 
$O(3)$ multi-critical point can be realized in the Heisenberg bilayer with 
dimer dilution, i.e., where adjacent spins on opposite layers are removed
together. This system is illustrated in Fig.~\ref{fig:bilayer}, and a 
schematic $T=0$ phase diagram is shown in Fig.~\ref{fig:phasediag}. 
In analogy with quantum critical points in clean 2D Heisenberg 
antiferromagnets \cite{chn,ye}, one can expect a finite-$T$ universal 
quantum critical scaling regime to extend to couplings well beyond the 
$T=0$ critical coupling $g^*$, possibly all the way to decoupled layers 
($g=0$). This quantum criticality could then be realized in layered 
antiferromagnets doped with non-magnetic impurities. It may already have 
been observed in  La$_{\rm 2}$Cu$_{\rm 1-x}$(Zn,Mg)$_{\rm x}$O$_{\rm 4}$,
for which recent neutron scattering experiments \cite{vajk} show a correlation
length divergence roughly consistent with the dynamic exponent $z\approx 1.3$ 
extracted here.

The clean $S=1/2$ bilayer Heisenberg model has been extensively studied in 
the past \cite{bilay1,bilay2}. It undergoes a quantum phase transition 
between an antiferromagnetic and a quantum disordered state as a function 
of the ratio $g=J_2/J_1$ of the inter- and intra-plane couplings. The
critical coupling $g_c \approx 2.52$ and the exponents are consistent with 
the expected \cite{chn} classical 3D Heisenberg universality class. If the
system is diluted by randomly removing single spins, the quantum phase
transition is destroyed because moments are induced around the ``holes''
in the gapped phase. These localized moments order antiferromagnetically 
for all $g$ \cite{wessel}. In order to circumvent this ``order from disorder''
phenomenon, inter-plane dimer dilution will be considered here. A dimer is 
not associated with moment formation and hence there is a spin gap for large 
$g$ at any dilution fraction $p$. When $g=0$ the system corresponds to two 
independent site-diluted Heisenberg layers. In that case, it was recently 
shown that the sublattice magnetization on large clusters at the percolation 
density scales to a non-zero value in the limit of infinite cluster size 
\cite{perc1,perc2}, implying a classical percolation transition. In 
the bilayer the long-range order on the percolating cluster can then be 
expected to survive up to a critical inter-layer coupling $g^* > 0$, leading
to the phase diagram shown in Fig.~\ref{fig:phasediag}. 

In order to extract the multi-critical coupling $g^*$, quantum Monte Carlo 
simulations similar to those discussed in Ref.~\onlinecite{perc2} were carried
out at the percolation point ($p^* \approx 0.4072538$ \cite{newman}). 
Two types of boundary conditions were used. 
In periodic $L \times L$ systems, dimers were removed with probability 
$p^*$ and the largest cluster of connected dimers was studied. These 
clusters have a varying number $N_1$ of dimers, with $\langle N_1\rangle 
\sim A L^d$, where the fractal dimension $d=91/48$ \cite{stauffer} and 
$A \approx 0.67$. Clusters were also constructed at fixed size 
$N$ without boundary imposed shape restrictions \cite{perc2}. In this case 
the corresponding length-scale is $R = N^{1/d}$. The two types of 
clusters will be referred to as $L \times L$ and fixed-$N$, respectively. A 
length $R=0.81\cdot L$ will sometimes be used for the $L \times L$ 
clusters, so that for a given $R$ the average size $\langle N_1 \rangle 
\approx R^{1/d}$ is the same as the size of the fixed-$N$ clusters.
The calculations were carried 
out using the stochastic series expansion method \cite{sse}. 
Temperatures sufficiently low to obtain 
ground state properties were used for $L \times L$ clusters with $L$ up to 
$64$ and fixed-$N$ clusters with $N$ up to $1024$. The results were averaged 
over a large number of dilution realizations; from $ > 10^3$ for the largest 
sizes to $> 10^5$ for smaller sizes.

\begin{figure}
\includegraphics[width=8.4cm]{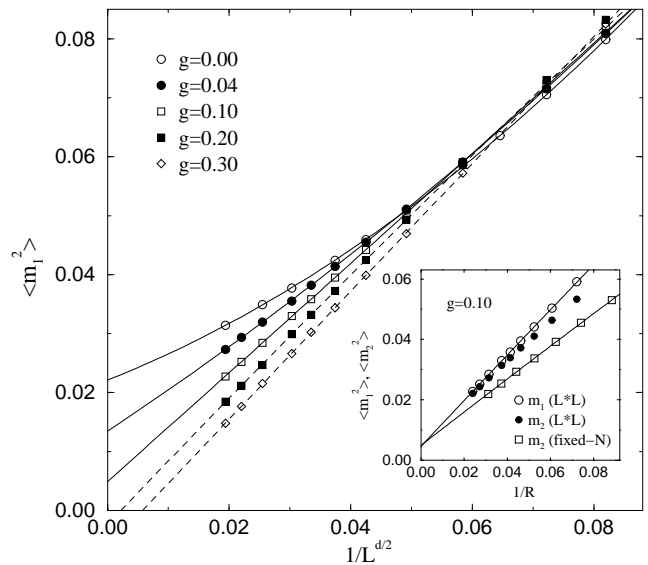}
\caption{Finite-size scaling of the single-plane sublattice magnetization 
on $L \times L$ clusters at the percolation density. In the inset, 
both the single-plane $(m_1)$ and full-system ($m_2$) definitions of the
sublattice magnetization on $L\times L$ lattices at $g=0.10$ are shown 
along with $m_2$ for the fixed-$N$ clusters.}
\label{fig:m1}
\end{figure}

For a cluster with $N_1$ dimers ($N_2=2N_1$ spins), single-plane ($a=1$) 
and full-system ($a=2$) staggered structure factors and susceptibilities 
are defined as
\begin{eqnarray}
S_a(\pi,\pi) & = & {1\over N_a} \sum\limits_{i,j=1}^{N_a}
P_{ij} \langle S^z_{i}S^z_{j} \rangle , \\
\chi_a(\pi,\pi) & = & {1\over N_a} \sum\limits_{i,j=1}^{N_a}
P_{ij} \int_0^\beta d\tau \langle S^z_{i}(\tau)S^z_{j}(0) \rangle ,
\end{eqnarray}
where $P_{ij}=1$ for sites $i,j$ on the same sublattice and $-1$ for sites
on different sublattices. Squared sublattice magnetizations are defined as
\cite{reger} $\langle m_a^2 \rangle = \langle 3S_a(\pi,\pi)/N_a \rangle$,
where $\langle \rangle$ indicates averaging over dilution realizations. The 
two definitions $\langle m_1^2\rangle$ and $\langle m_2^2\rangle$ should 
extrapolate to the same infinite-size value but the finite-size corrections
can be different. In clean systems, it is known that the leading size 
correction is $\sim N^{-1/2}$ \cite{huse} and this was found to be the case 
also for the diluted single layer \cite{perc2}. In Fig.~\ref{fig:m1}, 
$\langle m_1^2\rangle$ on $L \times L$ clusters is graphed versus $L^{-d/2} 
\sim \langle N_1 \rangle ^{-1/2}$ for several values of the coupling,
along with the previous $g=0$ results. Quadratic fits
are also shown. Interestingly, the sub-leading corrections become 
smaller, i.e., the behavior becomes more linear, as $g$ is increased 
from $0$. For $g=0.1$ the data for $L=20-64$ can be fitted to a 
purely linear form, which extrapolates to a non-zero value. In the inset of 
Fig.~\ref{fig:m1} it is shown that the definition $\langle m_2^2\rangle$ 
has more curvature but approaches $\langle m_1^2\rangle$ for the largest 
cluster sizes. Results for $\langle m_2^2\rangle$ on fixed-$N$ clusters are
also shown to extrapolate to the same infinite-size value, with an over-all
smaller size-correction. For $g \ge 0.2$ the extrapolations give negative
values, indicating that the sublattice magnetization vanishes (the fitted 
forms can here not be correct for very large $L$, as the asymptotic behavior 
has to be $~1/L^d$ if $\langle m_2^2\rangle=0$). Hence, the critical coupling 
$0.1 < g^* < 0.2$.

\begin{figure}
\includegraphics[width=7.5cm]{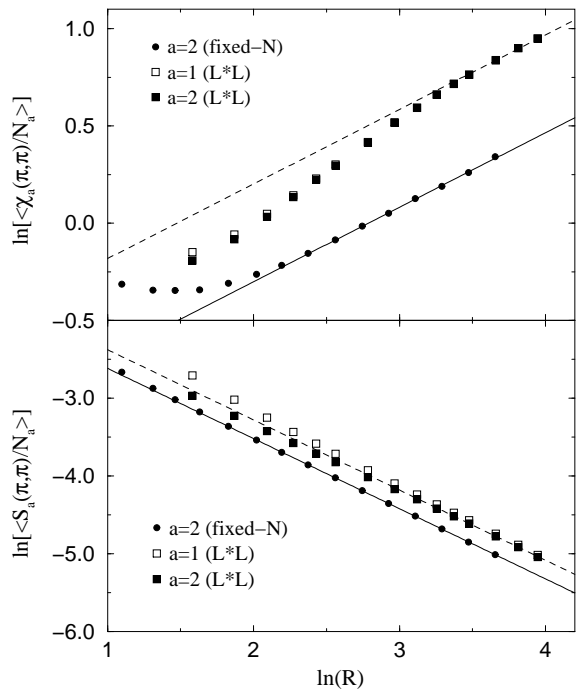}
\caption{Finite-size scaling of the staggered structure factors and 
susceptibilities. The solid lines are fits to results for the fixed-$N$
clusters; the dashed lines have the same slopes but are shifted to match
the $L\times L$ data for large $R$.}
\label{fig:sx}
\end{figure}

At ($g^*,p^*)$, both $S_a(\pi,\pi)$ and $\chi_a(\pi,\pi)$ should exhibit
power-law finite-size scaling. In clean systems the quantum critical scaling 
forms are $S \sim L^{1-\eta}$ and $\chi \sim L^{1+z-\eta}$, where 
$\eta \approx 0.03$ is the equal-time spin correlation function exponent and 
$z=1$ is the dynamic exponent. Since the size fluctuates in the case of the 
$L\times L$ clusters, the statistical errors in this case are reduced in the 
size-normalized quantities $\langle S_a/N_a\rangle$ and $\langle \chi_a/N_a
\rangle$, which therefore will be studied here. If the exponents
are defined according to
\begin{eqnarray}
\langle S_a(\pi,\pi)/N_a \rangle & \sim & R^{\gamma_S}, \label {spi} \\
\langle \chi_a(\pi,\pi)/N_a\rangle & \sim & R^{\gamma_\chi}, \label{xpi}
\end{eqnarray}
the dynamic exponent can be obtained from the difference; 
$z=\gamma_\chi - \gamma_S$. The best over-all scaling behavior is seen in 
$\langle S_2/N_2\rangle$ for the fixed-$N$ clusters. Based on this quantity, 
the multi-critical point is estimated to $g^* = 0.16 \pm 0.01$, and the 
exponent $\gamma_S = -0.90 \pm 0.01$. A log-log plot with data for both 
fixed-$N$ and $L \times L$ clusters at $g=0.16$ is shown in Fig.~\ref{fig:sx}. 
The $L \times L$ data have larger corrections to scaling but for the largest 
clusters the behavior is completely consistent with the exponent extracted 
from the fixed-$N$ data. The power-law scaling in $\langle \chi_a/N_a\rangle$ 
sets in at larger system sizes, and also in this case the behaviors seen 
for $L \times L$ and fixed-$N$ clusters are consistent with each other. 
The fixed-$N$ clusters again show a wider range of good scaling and give 
$\gamma_\chi = 0.38 \pm 0.03$. The dynamic exponent is hence $z = 1.28 
\pm 0.02$, where correlations between $\gamma_S$ and $\gamma_\chi$ have 
been taken into account in the error estimate. 

\begin{figure}
\includegraphics[width=7.0cm]{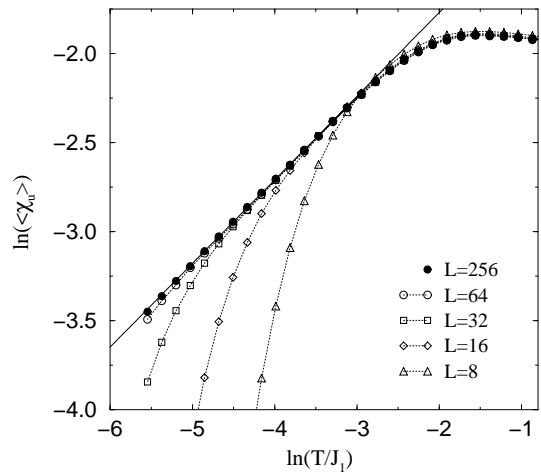}
\caption{Temperature dependence of the susceptibility at $g = 0.16$. 
Results for different system sizes are shown along with a linear fit to the 
$L=256$ data. The small deviations at the lowest temperatures indicate that 
$g^*$ is marginally below $0.16$.}
\label{fig:xu}
\end{figure}

According to hyper-scaling theory, the temperature dependence of the
uniform magnetic susceptibility $\chi_u$ should be governed by the
dynamic exponent \cite{fisher,ye}:
\begin{equation}
\langle \chi_u \rangle = {J_1\over T}
\left \langle {1\over N_2}
\left \langle \left (
\sum_{i=1}^{N_2} S^z_i \right )^2 \right\rangle\right\rangle
\sim T^{D/z-1}. 
\label{usus}
\end{equation}
Consistency with the $z$ extracted above from $T=0$ quantities can hence 
be tested in a non-trivial way. Finite-temperature calculations were carried 
out using sufficiently large $L \times L$ clusters to completely eliminate 
finite-size effects down to $T/J_1 = 1/256$. Results at $g=0.16$ are shown on 
a log-log plot in Fig.~\ref{fig:xu}. There is indeed a significant linear 
low-temperature regime, where the slope is $0.470 \pm 0.005$. Using $d=91/48$
for $D$ in Eq.~(\ref{usus}) then gives the dynamic exponent $z=1.29 
\pm 0.01$, in full agreement with the $T=0$ result.

\begin{figure}
\includegraphics[width=6.6cm]{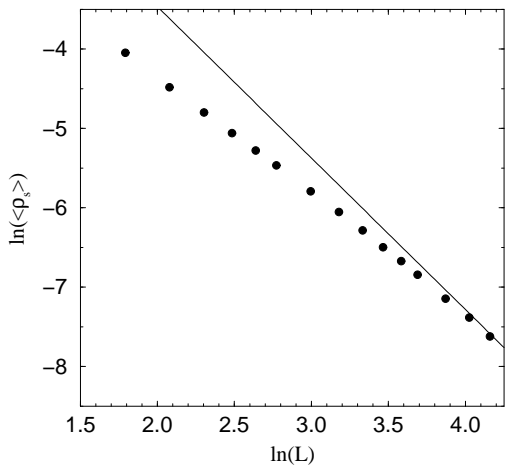}
\caption{Finite-size scaling of the $T=0$ spin stiffness at $g=0.16$. The
line shows the proposed asymptotic behavior.}
\label{fig:rho}
\end{figure}

Another important quantity is the spin stiffness $\rho_s$. It can be 
calculated in the simulations using the winding number fluctuations 
\cite{sse2} in systems with periodic boundary conditions (i.e., using
$L \times L$ clusters). Hyper-scaling predicts $\rho_s \sim L^{2-D-z}$ 
\cite{fisher,wallin} at a $T=0$ critical point. At the percolation point 
one can expect that $D$ should {\it not} be replaced by 
the fractal dimension $d$ of the percolating cluster, but by its backbone 
dimensionality $d_b$, which is significantly smaller ($d_b \approx  1.643$ 
\cite{grassberger}). This is because the spin currents wrapping around the
periodic clusters, in terms of which $\rho_s$ is evaluated, only flow through
the backbone. Furthermore, the above scaling form applies to systems in which 
the stiffness takes a finite constant value in the ordered phase. However, 
at $p^*$ in the single layer (and hence for all $g < g^*$) it scales as 
$\langle \rho_s \rangle \sim L^{-t/\nu}$, where $t$ is the conductivity 
exponent of percolation and $\nu=4/3$ is the percolation correlation length 
exponent \cite{harris2,perc2}. The ratio $t/\nu \approx 0.983$ according to 
recent simulations \cite{grassberger}. In order to account for the 
``geometric'' reduction, 
the following scaling form is tested here:
\begin{equation}
\langle \rho_s \rangle \sim L^{2-d_b-z-t/\nu}.
\label{rhoscale}
\end{equation}
Using the value extracted for $z$ above, the exponent 
$2- d_b - z - t/\nu \approx -1.92$. Fig.~\ref{fig:rho} shows data at $g=0.16$ 
along with this power-law scaling. There are clearly deviations, 
but it is also apparent that the numerical results are not yet in the 
asymptotic scaling regime. Significant corrections to the scaling law
$L^{-t/\nu}$ were also seen in the single-layer case \cite{perc2}. A definite
test of the conjectured form (\ref{rhoscale}) hence requires calculations for
larger system sizes.

To summarize the results, it has been shown that a multi-critical point at 
the percolation threshold is realized in the dimer-diluted Heisenberg 
bilayer at a critical inter-plane coupling $g^*=0.16 \pm 0.01$. The dynamic 
exponent $z$ was determined using both $T=0$ and $T>0$ quantities, with 
both calculations consistent with $z = 1.28 \pm 0.02$. Vajk and Greven have 
recently studied the same model at $T > 0$ \cite{vajk2}, with results 
in good agreement with those presented here.

In analogy with finite-temperature quantum criticality in clean 2D
antiferromagnets \cite{chn,ye}, there should be a significant universal
quantum critical regime controlled by the point $(p^*,g^*)$. A very
interesting question is then whether this universality could be observed 
even in a single diluted layer. The correlation length should diverge as 
$T^{-1/z}$ in the quantum critical regime \cite{chn,ye}. It is intriguing 
that this form with $z \approx 1.4$ has recently been observed at high 
temperatures, in simulations as well as in neutron scattering experiments 
on La$_{\rm 2}$Cu$_{\rm 1-p}$(Zn,Mg)$_{\rm p}$O$_{\rm 4}$ \cite{vajk}. 
However, since the experimental system at the percolation point corresponds 
to $g<g^*$ there should, in the absence of 3D effects, be a $T\to 0$ 
cross-over to a different dynamic exponent characterizing the line 
$(p=p^*,g<g^*)$. Although the perolating cluster is ordered, the exponential 
``renormalized classical'' behavior \cite{chn,chernyshev} cannot apply 
here because $\rho_s=0$. 

The $T=0$ dynamic exponent for $(p=p^*,g<g^*)$ is governed by the
properties of the percolating cluster. Since it is long-range ordered 
the structure factor exponent $\gamma_S=0$ in Eq.~(\ref{spi}). In a clean 
$D$-dimensional system with long-range order the staggered susceptibility 
diverges as $L^{2D}$ \cite{hasen}, and if this holds also at the percolation 
point it would imply $\gamma_\chi=d$ in Eq.~(\ref{xpi}) and hence $z(g<g^*)=d
=91/48$. This is in contrast to the transverse Ising model, where there is 
activated scaling, i.e., $z=\infty$ at the percolation transition 
\cite{senthil}. The bilayer simulations are consistent with $z(g \alt 0.1)=d$.
Closer to $g^*$ cross-over effects make it hard to verify this asymptotic 
behavior. 

It should be pointed out that the bilayer coupling used here to realize a 
multi-critical point $(p^*,g^*)$ is only one way to achieve this universality 
class. In Zn and Mg doped cuprates there may be interactions driving the 
system from $g=0$ closer to such a point in an extended parameter space. 
The fact that the sublattice magnetization measured experimentally 
\cite{vajk} falls significantly below the calculated curve \cite{perc2} as 
$p \to p^*$ indicates that such couplings indeed are present.

I would like to thank I. Affleck, A. Castro Neto, M. P. A. Fisher, M. Greven,
S. Sachdev, O. Sushkov, and O. Vajk for useful discussions and comments. 
This work was supported by the Academy of Finland (project 26175) and the
V\"ais\"al\"a Foundation. Some calculations were done on the Condor systems 
at the University of Wisconsin - Madison and the NCSA in Urbana, Illinois.

\null

\end{document}